\documentstyle[aps]{revtex}
\begin{document}

\title{QCD, STRONG CP AND AXIONS}

\author{R. D. Peccei}
\address{Department of Physics, University of California, Los Angeles,\\
         Los Angeles, CA  90095-1547}

\maketitle
\begin{abstract}

The physical origin of the strong CP problem in QCD, rooted in the 
structures of the vacuum of the standard model, is reviewed.  The chiral 
solution to this problem, with its accompanying axion, is explained and
various characteristics of axions are detailed.  Although visible axion
models are excluded experimentally, models with very light and very weakly 
coupled axions are still tenable.  The astrophysical and cosmological
implications of such axion models are discussed, along with ongoing
experimental attempts to detect such, so called, invisible axions.
\end{abstract}

%\narrowtext
%\twocolumn

\section*{The $U(1)_A$ Problem}

Quantum Chromodynamics (QCD) describes the strong interactions of hadrons in
terms of the interactions of their quark and gluon constituents.  The QCD
Lagrangian
\begin{equation}
{\cal{L}}_{\rm QCD} = -\sum_f \bar q_f\left(\gamma^\mu\frac{1}{i}D_\mu
+m_f\right) q_f - \frac{1}{4} G_a^{\mu\nu}G_{a\mu\nu}
\end{equation}
for $f$ flavors of quarks has a large global symmetry in the limit when
$m_f \to 0:~
G = U(f)_{\rm R}\times U(f)_{\rm L}$.
This symmetry
corresponds to the freedom of arbitrary chiral rotations of the $f$ flavor
of quarks into each other.  Because $m_u,m_d\ll\Lambda_{\rm QCD}$---with
$\Lambda_{\rm QCD}$ being the dynamical scale of the theory---in practice
only a chiral $U(2)_{\rm R}\times U(2)_{\rm L}$ symmetry is actually a very good
approximate global symmetry of the strong interactions.  This symmetry,
however, is not manifest in the spectrum of hadrons.  It is instructive to 
understand why this is so.

The $U(1)_{\rm V}$ subgroup of the $U(2)_{\rm R}\times U(2)_{\rm L}$ symmetry, corresponding to
vectorial (V = R+L) baryon number, is an exact symmetry of QCD, 
irrespective of the value of the quark masses.
In the limit of $m_u=m_d$  
a further $SU(2)_V$ subgroup
would be an exact (isospin) symmetry of QCD. 
Because the $u$- and $d$-quark masses are so light compared to 
$\Lambda_{\rm QCD}$, even with $m_u\not= m_d$
one has an approximately degenerate multiplet of hadrons
corresponding to this symmetry (e.g. the $\pi$-triplet and the doublet of
neutrons and protons).  However, the corresponding axial symmetries in
$U(2)_{\rm R}\times U(2)_{\rm L}$ ($A=R-L$) are not seen in the spectrum.
For instance, there are no parity doublets degenerate with the neutron and
proton.  This phenomena is understood because the $SU(2)_{\rm A}$ symmetry is 
not preserved by the QCD vacuum.  Indeed, as is appropriate for a spontaneously
broken global symmetry, there is instead in the QCD hadron spectrum an
approximate triplet of Nambu-Goldstone bosons---the pions---whose mass vanishes
in the limit that $m_u,m_d\to 0$.  Remarkably, however, there is no corresponding
pseudoscalar state with vanishing mass, in the same limit, corresponding to
the Nambu-Goldstone boson of a $U(1)_A$ symmetry.

The nature of the approximate $SU(2)_{\rm V}\times SU(2)_{\rm A}\times
U(1)_{\rm V}$ symmetry of the strong interactions was actually understood before the
advent of QCD\cite{GMDR}.  Although the anomalous features of the
$U(1)_{\rm A}$ symmetry were also pinpointed around the time of the
development of QCD as the theory of the strong interactions\cite{Weinberg},
the resolution of the $U(1)_{\rm A}$ problem required a better understanding
of the QCD vacuum.  In short, the reason why there are no approximate 
Nambu-Goldstone bosons associated with this Abelian chiral symmetry 
is that, as a result
of chiral anomaly\cite{ABJ}, this 
is really {\bf not} a quantum symmetry of QCD.
In turn, however, this more complete understanding
opens up another problem---the strong CP problem\cite{RDP}.  
Namely, the richer vacuum of
QCD, combined with the violation of CP in the weak interactions, allows for the
presence of an effective interaction
\begin{equation}
{\cal{L}}_{\rm strong~CP} = \bar\theta \frac{\alpha_s}{8\pi} G_{a\mu\nu}
\tilde G_a^{\mu\nu}
\end{equation}
which leads to an enormous neutron electric dipole moment $\left[d_n\sim
e\bar\theta (m_q/M_N^2)\right]$, unless the parameter $\bar\theta$ is 
tiny $(\bar\theta \leq 10^{-9})$.

In QCD, the dynamical formation of quark condensates, $\langle\bar uu\rangle=
\langle\bar dd\rangle\not= 0$, breaks the $U(2)_{\rm A}$ global symmetry.
As a result,
in the limit of vanishing quark masses, the pion triplet are the Nambu
Goldstone bosons associated with the spontaneously broken $SU(2)_{\rm A}$
current
$J^\mu_{5i} = \bar q\gamma^\mu\gamma_5 \frac{\tau_i}{2} q$,
where $q$ is the $(u,d)$ doublet of quark fields.  The existence of these
massless states produces a $q^2=0$ pole in the Green's functions of the
$SU(2)_{\rm A}$
currents with pseudoscalar quark densities.  One has
\begin{eqnarray}
M^\mu_{ij}(q) &=& \int d^4x e^{-iqx}
\langle T(J^\mu_{5i}(x),\bar q\gamma_5\tau_jq)\rangle  \nonumber \\
&\sim & \frac{\langle \bar qq\rangle\delta_{ij}q^\mu}{q^2} +
\hbox{non singular terms}~.
\end{eqnarray}
Because the $\langle\bar qq\rangle$ 
condensates also break the
$U(1)_{\rm A}$ global symmetry, 
naively one expects also a similar $q^2$ singularity
associated with the $U(1)_{\rm A}$ current, connected to a further 
Nambu-Goldstone boson--the $\eta$ in this 2-flavor discussion.  However, such an
approximate Nambu-Goldstone excitation cannot really exist, since if it did,
it would lead (in the limit of finite but small quark masses) to a state
essentially degenerate with the pions\cite{Weinberg}---in contradiction with
experiment.

The situation is a bit more complicated because the $U(1)_{\rm A}$ current
$J^\mu_5 = \bar q \gamma^\mu\gamma_5\frac{1}{2} q$
has a chiral anomaly\cite{ABJ}
\begin{equation}
\partial_\mu J^\mu_5 = \frac{\alpha_s}{4\pi} G_a^{\mu\nu}
\tilde G_{a\mu\nu}~.
\end{equation}
However, since the RHS of the above equation is a total divergence\cite{Bardeen}
\begin{eqnarray}
&& G_a^{\mu\nu}\tilde G_{a\mu\nu} = \partial_\mu K^\mu~; \nonumber \\
&& K^\mu = \epsilon^{\mu\alpha\beta\gamma}A_{a\alpha}
\left[G_{a\beta\gamma}-\frac{g_s}{3}f_{abc}A_{b\beta}A_{c\gamma}\right]~,
\end{eqnarray}
one can construct a conserved chiral current
\begin{equation}
\tilde J^\mu_5 = J^\mu_5 - \frac{\alpha_s}{4\pi} K^\mu~.
\end{equation}
As a result, in the massless quark limit, if 
$\langle\bar qq\rangle\not= 0$, one can show
that the Green's functions of the $\tilde J^\mu_5$ current have a $q^2=0$ singularity analogous to that of the chiral $SU(2)_{\rm A}$ currents:
\begin{eqnarray}
\tilde M^\mu(q) &=& \int d^4x e^{-iqx}
\langle T(\tilde J^\mu_5(x),~\bar q\gamma_5q)\rangle  \nonumber \\
&\sim &\frac{\langle\bar qq\rangle q^\mu}{q^2} +
\hbox{non-singular terms}~.
\end{eqnarray}
The $U(1)_{\rm A}$ problem, then, is why such a singularity is not manifested
physically in the spectrum of hadrons?

\section*{The QCD Vacuum}

The resolution of the $U(1)_{\rm A}$ problem came from a better understanding
of the structure of the QCD vacuum.  For gauge fields one can take the vacuum
state to be the state where the vector potential is either zero or is in a
gauge configuration equivalent to zero.  A proper study of these
configurations reveals that the structure of the vacuum state for non-Abelian
gauge theories (like QCD) is richer than expected.  Following Callan, Dashen,
and Gross\cite{CDG} it is convenient to study these theories in the
temporal gauge $A^o_a = 0$ and concentrate on an $SU(2)$ subgroup.  With
this gauge choice, the spatial gauge fields are time independent and under a
gauge transformation transform as
\begin{equation}
\frac{\tau_a}{2} A^i_a(\vec r) \equiv A^i(\vec r) \to
\Omega(\vec r) A^i(\vec r) \Omega(\vec r)^{-1} + \frac{i}{g_s}
\Omega(\vec r)\nabla^i\Omega^{-1}(\vec r)~. 
\end{equation}
Thus, in this gauge, vacuum configurations corresponds to a vanishing vector potential,
or to $\frac{i}{g_s}\Omega(\vec r)\nabla^i\Omega^{-1}(\vec r)$.

The rich vacuum structure of QCD results from the requirement that the gauge
transformation matrices $\Omega(\vec r)$ go to unity at spatial infinity.
Such a requirement maps the physical space onto the group space and this 
$S_3 \to S_3$ map splits $\Omega(\vec r)$ into different homotopy classes
$\{\Omega_n(\vec r)\}$, characterized by an integer winding number $n$ 
detailing 
how precisely $\Omega(\vec r)$ behaves as $\vec r\to\infty$:
\begin{equation}
\Omega_n(\vec r) \mathrel{\mathop\rightarrow_{\vec r\to \infty}}
e^{2\pi in}~.
\end{equation}
The winding number $n$ is related to the Jacobian of the $S_3\to S_3$
transformation
\begin{equation}
n= \frac{ig_s^3}{24\pi^2} \int d^3r {\rm Tr}~\epsilon_{ijk} A^i_n(\vec r)
A^j_n(\vec r) A^k_n(\vec r)~,
\end{equation}
with $A^i_n(\vec r)$ being the pure gauge field corresponding to the gauge
transformation matrix $\Omega_n(\vec r)$.

Because one can construct the $n$-gauge transformation matrix $\Omega_n(\vec r)$
by compounding $n$ times $\Omega_1(\vec r)$, it follows that the vacuum
state corresponding to $A_n(\vec r)$ is not really gauge invariant.  Indeed,
the action of the gauge transformation matrix $\Omega_1$ on an $n$-vacuum
state gives an $(n+1)$-vacuum state
\begin{equation}
\Omega_1|n\rangle = |n+1\rangle~.
\end{equation}
Nevertheless, one can construct a gauge-invariant vacuum state---the, so 
called, $\theta$-vacuum--by superposing these $n$-vacua\cite{CDG}
\begin{equation}
|\theta\rangle = \sum_n e^{-in\theta}|n\rangle~.
\end{equation}
Indeed, it is easy to check that
$\Omega_1|\theta\rangle = e^{i\theta}|\theta\rangle$.

Because of this more complex vacuum structure, the vacuum functional
for the theory splits into distinct sectors.
\begin{eqnarray}
_+\langle\theta|\theta\rangle_- &=& \sum_{m,n} e^{im\theta}
e^{-in\theta}{}_+\langle m|n\rangle_- \\
&=& \sum_\nu e^{i\nu\theta} \left[\sum_n
{}_+\langle n+\nu|n\rangle_-\right]~.
\end{eqnarray}
This superposition of transition amplitudes from $t=-\infty$ to $t=+\infty$
with fixed $n$-vacuum difference $\nu$, weighted by different phases
$e^{i\nu\theta}$ introduces an additional intrinsic source of CP violation.
By using Eqs. (5) and (10),
one can show\cite{RDP} that the integer $\nu$ has a gauge invariant meaning
\begin{equation}
\nu = \frac{\alpha_s}{8\pi} \int d^4x G_{a\mu\nu} \tilde G_a^{\mu\nu}~.
\end{equation}
Thus, writing for the vacuum functional the usual path integral formula
\begin{equation}
_+\langle\theta|\theta\rangle_-
= \sum_\nu \int 
\delta A^\mu e^{iS_{\rm eff} [A]} 
\delta \left[\nu - \frac{\alpha_s}{8\pi} \int d^4x G_a^{\mu\nu}
\tilde G_{a\mu\nu}\right]~,
\end{equation}
one can re-interpret the $\theta$ term in terms of an effective action
\begin{equation}
S_{\rm eff} = S[A] + \theta \frac{\alpha_s}{8\pi} \int d^4x
G_a^{\mu\nu} \tilde G_{a\mu\nu}~.
\end{equation}
This additional term violates P and CP, since it corresponds to an
$\vec E_a\cdot \vec B_a$ interaction of the color fields.

Perturbation theory is connected to the $\nu=0$ sectors where the $G\tilde G$
term vanishes.  The effects of the $\nu\not= 0$ sectors thus necessarily are
nonperturbative.  These contributions are naturally selected by the 
connection of $G\tilde G$ to the chiral anomaly.  For $n_f$ flavors, the axial
current $J^\mu_5$ has an anomaly
\begin{equation}
\partial_\mu J^\mu_5 = n_f \frac{\alpha_s}{8\pi} G_{a\mu\nu}\tilde G_a^{\mu\nu}
\end{equation}
Thus a chirality change $\Delta Q_5$ is related to $\nu$ as:
\begin{equation}
\Delta Q_5 = \int d^4x \partial_\mu J^\mu_5 = n_f
\frac{\alpha_s}{8\pi} \int d^4x G_a^{\mu\nu} \tilde G_{a\mu\nu} = n_f\nu
\end{equation}
The solution to the $U(1)_{\rm A}$ problem is connected to the chirality
breakdown in QCD in the $\nu\not= 0$ sectors.  If these sectors are included---as
they should---then $U(1)_ {\rm A}$ is {\bf never} a symmetry.

For $n_f=2$, the conserved current discussed earlier $\tilde J_5^\mu$
is not gauge invariant for transformations with non-trivial winding number.  
One can show\cite{RDP} that the associated chiral charge
\begin{equation}
\tilde Q_5 = \int d^3x \tilde J^o_5
\end{equation}
transforms as
\begin{equation}
\Omega_1 \tilde Q_5\Omega_1^{-1} = \tilde Q_5 + 2
\end{equation}
when one worries about $\nu\not= 0$ sectors. 't Hooft showed that 
it is this lack
of gauge invariance which is crucial to resolve the $U(1)_{\rm A}$ 
problem\cite{'t Hooft}. 
Although Green's functions of the
$\tilde J^\mu_5$ currents contain $1/q^2$ singularities, these singularities
are cancelled for the corresponding Green's functions of the gauge invariant
currents $J^\mu_5$.  The Green's function for this current with 
a pseudoscalar density obeys the equation
\begin{equation}
M^\mu(q) = \tilde M^\mu(q) + \frac{\alpha_s}{4\pi} \int d^4x e^{-iqx}
\langle T(K^\mu(x),~ \bar q\gamma_5q)\rangle
\end{equation}
Although the first term above is singular at $q^2=0$, when one considers the
effects of $\nu\not= 0$ sectors, so is the second term.
These singularities
{\bf cancel} for the Green's function of the physical current $J^\mu_5$\cite
{'t Hooft}.  Because of Eq. (19), the $U(1)_{\rm A}$ symmetry is really not a symmetry at all.  Therefore, not surprisingly, there are really 
no physical Nambu Goldstone bosons  
associated with this pseudo symmetry!

\section*{The Strong CP Problem}

The effective interaction (17) is not the only new source of CP violation
arising from the more complex structure of the QCD vacuum.  It is augmented
by an analogous term coming from the electroweak sector of the theory.
The origin of this additional interaction is as follows.  In general, the
mass matrix of quarks which emerges from the spontaneous breakdown of the
electroweak gauge symmetry is neither Hermitian nor diagonal:
\begin{equation}
{\cal{L}}_{\rm mass} = -\bar q_{{\rm R}_i}M_{ij}q_{{\rm L}_j}-
\bar q_{{\rm L}_i}(M^\dagger)_{ij} q_{{\rm R}_j}~.
\end{equation}
This matrix can be diagonalized by separate unitary transformations of the
chiral quark fields.  These transformations encompass a chiral
$U(1)_{\rm A}$ transformation
\begin{equation}
q_{\rm R}\to e^{i\alpha/2}q_{\rm R}~; ~~~
q_{\rm L} \to e^{-i\alpha/2}q_{\rm L}
\end{equation}
with $\alpha=\frac{1}{n_f}~\hbox{Arg det}~M$.  Such chiral transformations,
in effect, change the vacuum angle\cite{JR}.  That this is so follows
from Eq. (21).  Using this equation (replacing the factor of 2 by the number
of flavors $n_f$), it is easy to see that a chiral $U(1)_{\rm A}$ rotation
on the $\theta$ vacuum shifts the vacuum angle by Arg det $M$\cite{RDP}:
\begin{equation}
e^{i\alpha\tilde Q_5}|\theta\rangle =
|\theta + n_f\alpha\rangle = |\theta + \hbox{Arg det}~M\rangle~.
\end{equation}
Hence, in the full theory, the effective CP-violating interaction ensuing 
from the more complex structure of the QCD vacuum is that given in Eq. (2),
with
\begin{equation}
\bar\theta = \theta + \hbox{Arg det}~M~.
\end{equation}

The strong CP problem is really why the combination of QCD and electroweak
parameters which make up $\bar\theta$ should be so small.  As we remarked 
upon earlier, the effective interaction (2) gives a direct contribution to the
electric dipole moment of the neutron and the strong experimental bound on
$d_n$ requires that $\bar\theta\leq 10^{-9}$\cite{BC}.  In principle, because
$\bar\theta$ is a free parameter in the theory, any value of $\bar\theta$ is
equally likely.  However, one would like to obtain an understanding
from some underlying physics of why
this number is so small.

\section*{The Chiral Solution to the Strong CP Problem}

Almost twenty years ago, Helen Quinn and I\cite{PQ} suggested a dynamical
solution to the strong CP problem.  What we postulated was that the
{\bf full} Lagrangian of the standard model was invariant under an additional
global chiral $U(1)$ symmetry.  If this $U(1)_{\rm PQ}$ symmetry existed, 
and it were
exact, the strong CP problem would be trivially solved since $\bar\theta$ 
could be set
to zero through such a chiral transformation (cf. Eq. (25)).  However,
physically such a symmetry cannot be exact.  Nevertheless, what Quinn and 
I\cite{PQ} showed was that, even 
if $U(1)_{\rm PQ}$ is spontaneously broken, 
the parameter $\bar\theta$ is
dynamically driven to zero.  However, in this case there is an associated 
pseudo Nambu-Goldstone boson in the theory, the axion\cite{WW}.\footnote{The axion is not massless because the chiral $U(1)_{\rm PQ}$ symmetry is anomalous.
As a result, the axion gets a mass of order $m_a\sim \Lambda^2_{\rm QCD}/f$,
where $f$ is the scale associated with the breakdown of the $U(1)_{\rm PQ}$
symmetry.}

It is useful to understand schematically how a global $U(1)_{\rm PQ}$ symmetry solves
the strong CP problem\cite{RDP}.  Basically, what happens is that by
incorporating this symmetry in the theory one replaces the static CP violating
parameter $\bar\theta$ by the dynamical (CP conserving) interactions of the
axion field $a(x)$.  Because the axion field is the Nambu-Goldstone boson
associated with the spontaneously broken $U(1)_{\rm PQ}$ symmetry, this field
translates under a $U(1)_{\rm PQ}$ transformation.  If $\alpha$ is the phase
parameter of this transformation and $f$ is the scale associated with the
breakdown of the symmetry, one has
\begin{equation}
a(x) \mathrel{\mathop \rightarrow_{U(1)_{\rm PQ}}} a(x) + \alpha f~.
\end{equation}
It follows, therefore, that if the effective Lagrangian describing the full
theory is to be $U(1)_{\rm PQ}$ invariant, the axion field must only enter
derivatively coupled.  This statement would be exactly true if the
$U(1)_{\rm PQ}$ symmetry were not anomalous.  Because of the chiral anomaly,
however, ${\cal{L}}_{\rm eff}$ must also have a term in which the axion field
couples directly to the gluon density $G\tilde G$, so as to guarantee that
$J^\mu_{\rm PQ}$ has the correct QCD chiral anomaly.

The above considerations fix the form of the effective standard model
Lagrangian, if it is augmented by an extra $U(1)_{\rm PQ}$ global
symmetry as we suggested\cite{PQ}.  Focusing only on the
extra terms involving the axion field one has
\begin{eqnarray}
{\cal{L}}^{\rm eff}_{\rm SM} = {\cal{L}}_{\rm SM} &+& \bar\theta
\frac{\alpha_s}{8\pi} G_a^{\mu\nu} \tilde G_{a\mu\nu} - \frac{1}{2}
\partial_\mu a\partial^\mu a  \nonumber \\
&+& {\cal{L}}_{\rm int.} [\partial^\mu a/f;\psi] +
\frac{a}{f} \xi \frac{\alpha_s}{8\pi} G_a^{\mu\nu} \tilde G_{a\mu\nu}~.
\end{eqnarray}
In the above $\psi$ stands for any field in the theory and $\xi$ is a 
model-dependent parameter associated with the chiral anomaly of the
$U(1)_{\rm PQ}$ current
\begin{equation}
\partial_\mu J^\mu_{\rm PQ} = \xi \frac{\alpha_s}{8\pi} G_a^{\mu\nu}
\tilde G_{a\mu\nu}~.
\end{equation}
The presence of the last term in (28) provides an effective potential for the
axion field.  Thus its vacuum expectation is no longer arbitrary.  Indeed the
minimum of this potential determines the axion field VEV $\langle a\rangle$.
One has
\begin{equation}
\left\langle \frac{\partial V_{\rm eff}}{\partial a}\right\rangle = -
\frac{\xi}{f} \frac{\alpha_s}{8\pi}
\langle G_a^{\mu\nu} \tilde G_{a\mu\nu}\rangle \bigg |_{\langle a\rangle}
=0~.
\end{equation}
What Quinn and I showed\cite{PQ} is that the periodicity of the pseudoscalar
density expectation value $\langle G\tilde G\rangle$ in the relevant
$\theta$-parameter, $\bar\theta + \frac{\langle a\rangle}{f} \xi$, forces
the axion VEV to take the value
\begin{equation}
\langle a\rangle = -\frac{f}{\xi} \bar\theta~.
\end{equation}
This solves the strong CP problem, since ${\cal{L}}_{\rm SM}^{\rm eff}$, when
expressed in terms of the physical axion field, $a_{\rm phys} = a-\langle a\rangle$, no longer contains the CP violating $\bar\theta G\tilde G$ term.
Furthermore, expanding $V_{\rm eff}$ at its minimum, one sees that the
axion itself gets a mass:
\begin{equation}
m_a^2 = \left\langle\frac{\partial^2V_{\rm eff}}{\partial a^2}\right\rangle =
-\frac{\xi}{f}\frac{\alpha_s}{8\pi}\frac{\partial}{\partial a}
\langle G_a^{\mu\nu} \tilde G_{a\mu\nu}\rangle \bigg |_{\langle a\rangle}~.
\end{equation}
Therefore, the standard model
 with an additional $U(1)_{\rm PQ}$ symmetry no 
longer has a dangerous CP violating interaction.  Instead, it contains additional
interactions of a massive axion field both with matter and gluons
characterized by a scale $f$:
\begin{eqnarray}
{\cal{L}}_{\rm SM}^{\rm eff} = {\cal{L}}_{\rm SM} &+& {\cal{L}}_{\rm int.}
[\partial^\mu a_{\rm phys.}/f;\psi] - \frac{1}{2} \partial^\mu a_{\rm phys.}
\partial_\mu a_{\rm phys.}  \nonumber  \\
&-& \frac{1}{2} m_a^2 a^2_{\rm phys.} + \frac{a_{\rm phys.}}{f} \xi
\frac{\alpha_s}{8\pi} G_a^{\mu\nu} \tilde G_{a\mu\nu}~.
\end{eqnarray}

\section*{Axion Dynamics}

In practical applications, it is more convenient to use the freedom of 
$U(1)_{\rm PQ}$ transformations to replace the last term in Eq. (33) in favor
of effective interactions of axions with the light pseudoscalar mesons
($\pi$ and $\eta$) of QCD\cite{BPY}.  Although the properties of axions are 
model-dependent, this dependence can be isolated in a few numerical
coefficients related to the axion mass, its coupling to two photons and the
mixing of axions with the neutral pion and the eta.  In terms of the pion decay
constant $f_\pi \simeq 92~{\rm MeV}$ and the $SU(2)\times U(1)$ VEV
$v\simeq 250~{\rm GeV}$, it proves convenient to define a ``standard" axion mass
parameter
\begin{equation}
m_a^{\rm st} = \frac{m_\pi f_\pi}{v} 
\frac{\sqrt{m_um_d}}{(m_u+md)} \simeq 25~{\rm KeV}~.
\end{equation}
Then, for all models, one can characterize the mass of the axion and
its $\pi$ and $\eta$ couplings by
\begin{equation}
m_a = \lambda_m m_a^{\rm st} \left(\frac{v}{f}\right)
\end{equation}
and
\begin{equation}
\xi_{a\pi} = \lambda_3 \frac{f_\pi}{f}~; ~~~
\xi_{a\eta} = \lambda_o \frac{f_\pi}{f}~,
\end{equation}
with $\lambda_m,~\lambda_3$, and $\lambda_o$ model parameters of
$O(1)$.\footnote{In certain models, it is possible for $\lambda_3$ or
$\lambda_o$ to vanish identically.}  Similarly, one can characterize the
effective couplings of axions to two photons by the interaction
\begin{equation}
{\cal{L}}_{a\gamma\gamma} = \frac{\alpha}{4\pi}
K_{a\gamma\gamma} \frac{a_{\rm phys.}}{f} 
F^{\mu\nu} \tilde F_{\mu\nu}~,
\end{equation}
where $K_{a\gamma\gamma}$ is again a model-dependent parameter of $O(1)$.
It is clear that if $f_\pi\ll f$, the axion is both very light and very weakly 
coupled!

It is informative to sketch the derivation of the above formulas in the
original Peccei-Quinn model\cite{PQ}.  Here the $U(1)_{\rm PQ}$
symmetry is introduced
in the standard model by having two Higgs doublets:
\begin{equation}
{\cal{L}}_{\rm Yukawa} = \Gamma^u_{ij} \bar Q_{{\rm L}_i} \Phi_1 
u_{{\rm R}_j} +
\Gamma^d_{ij}\bar Q_{{\rm L}_i}\Phi_2 d_{{\rm R}_j} + {\rm h.c.}
\end{equation}
The presence of $\Phi_1$ and $\Phi_2$ allows ${\cal{L}}_{\rm Yukawa}$
to be invariant under independent rotations of $u_{\rm R}$ and 
$d_{\rm R}$--the desired chiral $U(1)_{\rm PQ}$ transformation.  The axion
in this model is the common phase field of $\Phi_1$ and $\Phi_2$
orthogonal to weak hypercharge.  If $x=v_2/v_1$ is the ratio of the Higgs VEV 
and $f = \sqrt{v_1^2+v_2^2} = (\sqrt{2} G_F)= v \simeq 
250~{\rm GeV}$,
then it is easy to isolate the axion content in $\Phi_1$ and
$\Phi_2$ as
\begin{equation}
\Phi_1 = \frac{v_1}{\sqrt{2}}
\left[
\begin{array}{c}
1 \\ 0
\end{array}
\right]
e^{ixa/f}~; ~~~~
\Phi_2 = \frac{v_2}{\sqrt{2}}
\left[
\begin{array}{c}
0 \\ 1
\end{array}
\right]
e^{ia/xf}~.
\end{equation}
Using Eq. (27), to guarantee invariance of ${\cal{L}}_{\rm Yukawa}$,
$u_{\rm R}$ and $d_{\rm R}$ transform under a $U(1)_{\rm PQ}$ transformation
as
\begin{equation}
u_{\rm R}\to e^{-i\alpha x}u_{\rm R}~; ~~~
d_{\rm R} \to e^{-i\alpha/x} d_{\rm R}~.
\end{equation}
Whence, it is easy to see that the current associated with 
the $U(1)_{\rm PQ}$ symmetry is:\footnote{Here I assumed
that the charged leptons $\ell_{\rm R}$ have a Yukawa coupling involving
$\Phi_2$.}
\begin{eqnarray}
J^\mu_{\rm PQ} = -f\partial^\mu a &+& x\sum^{N_g}_{i=1}
\bar u_{{\rm R}_i} \gamma^\mu u_{{\rm R}_i} + \frac{1}{x}
\sum^{N_g}_{i=1} \bar d_{{\rm R}_i} \gamma^\mu d_{{\rm R}_i} \nonumber \\
&+& \frac{1}{x} \sum^{N_g}_{i=1} \bar\ell_{{\rm R}_i} \gamma^\mu
\ell_{{\rm R}_i}~.
\end{eqnarray}
This $U(1)_{\rm PQ}$ current has both a QCD and an electromagnetic anomaly
\begin{equation}
\partial_\mu J^\mu_{\rm PQ} = \xi \frac{\alpha_s}{8\pi} G_{a\mu\nu}
\tilde G_a^{\mu\nu} + \xi_\gamma\frac{\alpha}{4\pi} F_{\mu\nu}
\tilde F^{\mu\nu}~,
\end{equation}
where\cite{RDP}
\begin{equation}
\xi = N_g \left(x + \frac{1}{x}\right)~; ~~~
\xi_\gamma = N_g \frac{4}{3} \left(x + \frac{1}{x}\right)~.
\end{equation}

The QCD anomaly renders a bit more difficult the calculation of the
effective interactions of axions with light hadrons\cite{BT}.  Perhaps
the simplest way to proceed is by using an effective Lagrangian 
technique\cite{BPY} to describe the interactions of pions and the $\eta$,
both among themselves and with axions.  Since $\pi$ and $\eta$ are the
Nambu-Goldstone bosons associated with the approximate global
$U(2)_{\rm R}\times U(2)_{\rm L}$ symmetry of QCD, their interactions are
described by an effective chiral Lagrangian
\begin{equation}
{\cal{L}}_{\rm chiral} = -\frac{1}{4} f_\pi^2 {\rm Tr}~
\partial_\mu \Sigma^\dagger \partial^\mu\Sigma
\end{equation}
with
\begin{equation}
\Sigma = \exp\frac{i}{f_\pi} [\vec\tau\cdot\vec\pi + \eta]~.
\end{equation}
This Lagrangian must be augmented by a mass breaking term that mimics the
Yukawa interactions (38) and thus involves the axion field, along with an
axion kinetic energy term
\begin{equation}
{\cal{L}}_{\rm axion} = -\frac{1}{2} f_\pi^2 m_\pi^2 ~{\rm Tr}
[\Sigma AM+M^\dagger A^\dagger \Sigma^\dagger] - \frac{1}{2}
\partial_\mu a\partial^\mu a~,
\end{equation}
where
\begin{equation}
A=
\left[
\begin{array}{cc}
e^{-ixa/f} & 0 \\  0 & e^{-ia/xf}
\end{array}
\right]~; ~~~
M=
\left[
\begin{array}{cc}
\frac{m_u}{m_u+m_d} & 0 \\
0 & \frac{m_d}{m_u+m_d}
\end{array}
\right]~.
\end{equation}
The presence of $A$ guarantees the $U(1)_{\rm PQ}$ invariance of 
${\cal{L}}_{\rm axion}$ since under $U(1)_{\rm PQ}$, to mimic the quark
transformations,
\begin{equation}
\Sigma\to\Sigma
\left[
\begin{array}{cc}
e^{ix\alpha} & 0 \\
0 & e^{i\alpha/x}
\end{array}
\right]~.
\end{equation}
The anomaly interactions, which break the above 
$U(2)\times U(2)\times U(1)_{\rm PQ}$
symmetry through the coupling of gluons to axions and the $\eta$, serve to give
an effective mass term to the field combination which couples to
$G\tilde G$.  One has:\footnote{The factor of $(N_g-1)$ in 
${\cal{L}}_{\rm anomaly}$ enters since here one must include only the
contribution of the heavy quarks in the axion-gluon interactions, as the light
quark interactions are explicitly accounted for in ${\cal{L}}_{\rm axion}$.}
\begin{equation}
{\cal{L}}_{\rm anomaly} = -\frac{1}{2} m^2
\left[\eta + \frac{f_\pi}{f} \frac{(N_g-1)}{2}
\left(x + \frac{1}{x}\right) a\right]^2~,
\end{equation}
with $m^2\simeq m^2_\eta \gg m^2_\pi$.

The quadratic terms in ${\cal{L}}_{\rm axion}$ along with 
${\cal{L}}_{\rm anomaly}$, when diagonalized, allows one immediately
to compute the axion mass in the model, as well as the axion mixings with
$\pi^o$ and $\eta$
\begin{equation}
a\simeq a_{\rm phys.} - \xi_{a\pi}\pi^o_{\rm phys.} -
\xi_{a\eta}\eta_{\rm phys.}~.
\end{equation}
A simple calculation yields for the Peccei-Quinn model\cite{PQ}:
\begin{eqnarray}
\lambda_m &=& N_g\left(x + \frac{1}{x}\right)~; \nonumber \\
\lambda_3 &=& \frac{1}{2}\left[\left(x-\frac{1}{x}\right) -
N_g\left(x + \frac{1}{x}\right)
\frac{(m_d-m_u)}{(m_d+m_u)}\right]~; \nonumber \\
\lambda_o &=& \frac{1}{2}(1-N_g)\left(x + \frac{1}{x}\right)~.
\end{eqnarray}
Using these results, one can deduce readily the effective axion to two-photon
coupling.  The electromagnetic anomaly of the $\pi^o$ and $\eta$ fields
\begin{equation}
{\cal{L}}_{\pi,\eta} = \frac{\alpha}{4\pi}
\left[\frac{\pi^o}{f_\pi} + \frac{5}{3}\frac{\eta}{f_\pi}\right]
F_{\mu\nu}\tilde F^{\mu\nu}~,
\end{equation}
through the mixing with axions, gives an effective $a\gamma\gamma$ coupling coming from
the light quark sector.  To this coupling, one must add the heavy quark
contributiion.  This is given by Eq. (42), except that for the parameter
$\xi_\gamma$ one should replace $N_g$ by $N_g-1$. Adding these contributions
together yields finally, for the model,
\begin{equation}
K_{a\gamma\gamma} = N_g\left(x + \frac{1}{x}\right)
\frac{m_u}{m_u+m_d}~.
\end{equation}

\section*{The Demise of Visible Axion Models}

By changing the detailed way in which the Higgs fields $\Phi_1$ and
$\Phi_2$ couple to the quarks and leptons, one can obtain a variety of axion
models\cite{KP}.  All of these variant models, as well as the original
Peccei-Quinn model, have been ruled out experimentally.  I will not discuss
in detail here the different evidence against these weak scale ($f=v$)
axions, as this is reviewed in depth in Ref.\cite{RDP}.  I will, however,
give a few examples to give a flavor both of the models and of the
experiments used to rule these axions out.

If $f=v$, then the resulting axions are quite light, since
$m_a=25\lambda_m~{\rm KeV}$.  Unless $\lambda_m \gg 1$, then $m_a<2m_e$ and
these axions are also very long lived, since they can only decay via the
$a\to 2\gamma$ process.  These light, long lived, weak scale axions are
ruled out by the non-observation of the process $K^+\to \pi^+a$.  One
predicts\cite{BPY}
\begin{equation}
BR(K^+\to\pi^+a)\simeq 3\times 10^{-5} \lambda_o^2
\end{equation}
while, experimentally, from KEK there is a bound\cite{Asano}:
\begin{equation}
BR(K^+\to\pi^+~\hbox{nothing}) < 3.8\times 10^{-8}~.
\end{equation}
One can avoid this bound in models where $\lambda_o=0$.\footnote{This requires
that, effectively, only the first generation of quarks $(N_g=1)$ feel the
$U(1)_{\rm PQ}$ symmetry.}  However, experiments looking for axion production
via nuclear de-excitation in reactors are sensitive to both $\lambda_o$ and
$\lambda_3$.  The absence of a tell-tale $a\to 2\gamma$ signal downstream,
although less stringent than the $K^+\to\pi^+a$ process, then serves to rule out
this variant\cite{Zehadev}.

Weak scale axions can be short-lived if $m_a>2m_e$.  For this to happen,
$\lambda_m$ must be large and this necessitates either $x$ or $x^{-1}$ to be
large.  This possibility runs into difficulty with experiments looking for
axions in quarkonium decays, $Q\bar Q\to a\gamma$.  If the $U(1)_{\rm PQ}$
assignments of the Higgs fields are as in the original Peccei-Quinn 
model\cite{PQ}, then the rate for $\psi\to a\gamma$ is proportional to $x^2$
and that for $\Upsilon\to a\gamma$ is proportional to $x^{-2}$.  The present
bounds on these processes\cite{PQ} are only consistent with values of
$x\sim O(1)$.  Thus short-lived axion models to be viable must have
variant quark couplings,\cite{KP} such that both the $\psi\to a\gamma$ and
$\Upsilon\to a\gamma$ rates are either proportional to $x^2$ or $x^{-2}$.

These variant models are ruled out by a combination of other experiments.\cite
{BPY}  The process $\pi^+\to e^+e^-e^+\nu_e$ observed at SIN\cite{SIN} with
a branching ratio of $O(10^{-9})$ bounds the parameter $\lambda_3$ in these
models, since the decay chain $\pi^+\to ae^+\nu_e$ followed by
$a\to e^+e^-$ would contribute to the signal.  One finds\cite{BPY} from these
considerations that
$\lambda_3\leq 0.2$.  On the other hand, the $I=0$ to $I=0$ M1 
decay of the 3.58 MeV excited state of $^{10}B$ to the ground state\cite{Calprice} provides a
bound on the isoscalar mixing parameter $\lambda_o$\cite{BPY},
$\lambda_o\leq 2$.  Although there are models in which one can satisfy either
one, or the other, of these two constraints, one cannot satisfy both because
these parameters are themselves constrained by the relation\cite{BPY}
\begin{equation}
|\lambda_3-\lambda_o|\simeq \frac{m_a}{m_\pi} \frac{v}{f_\pi}
\sqrt{\frac{m_u}{m_d}}  
\geq 15~.
\end{equation}

\section*{Invisible Axion Models}

Although it was a sensible assumption to suppose that the $U(1)_{\rm PQ}$
breaking scale $f$ was the same as the weak scale $v$, this is not
necessary.  The dynamical adjustment of the strong CP angle
$\bar\theta\to 0$ works for any scale $f$.  If $f\gg v$, the resulting axions
are very light $(m_a\sim 1/f)$, very weakly coupled ($\hbox{coupling}\sim
1/f$) and very long lived $\left(\tau(a\to 2\gamma)\sim f^5\right)$.  Thus
these axions are, apparently, invisible.

Because $f\gg v$ by assumption, in invisible axion models the
$U(1)_{\rm PQ}$ symmetry must be broken by an $SU(2)\times U(1)$ singlet
VEV.  Hence these models all introduce an $SU(2)\times U(1)$ singlet complex
scalar field $\sigma$.  The invisible axion is then, essentially, the phase
of $\sigma$.  Hence, concentrating again only on the axion degrees of 
freedom, one can write
\begin{equation}
\sigma = \frac{f}{\sqrt{2}} e^{ia/f}~.
\end{equation}

Broadly speaking, one can classify invisible axion models into two types
depending on whether or not they have direct couplings to leptons.  The,
so called, KSVZ axions\cite{KSVZ} are hadronic axions with only induced
coupling to leptons.  The, so called, DFSZ axions\cite{DFSZ}, on the other
hand arise in models where axions naturally couple to leptons already at
tree level.  I describe below these two types of invisible axions in some
more detail.

\subsection*{KSVZ Axions}

In these models one assumes that the ordinary quarks and leptons are PQ
singlets.  The $SU(2)\times U(1)$ singlet field $\sigma$, however, interacts
with some new heavy quarks $X$ which carry $U(1)_{\rm PQ}$ charge\cite{KSVZ}
via the Yukawa interaction:
\begin{equation}
{\cal{L}}_{\rm KSVZ} = -h\bar X_{\rm L} \sigma X_{\rm R} - h^*\bar X_{\rm R}
\sigma^\dagger X_{\rm L}~.
\end{equation}
The interactions of the KSVZ axions with ordinary quarks arises as a result
of the chiral anomaly which induces a coupling
\begin{equation}
{\cal{L}}_{\rm anomaly} = \frac{a}{f}
\left[\frac{\alpha_s}{8\pi} G_a^{\mu\nu} \tilde G_{a\mu\nu} +
3q_X^2 \frac{\alpha_s}{4\pi} F_{\mu\nu} \tilde F^{\mu\nu}\right]~,
\end{equation}
where $q_X$ is the electric charge of the heavy quarks $X$.  Calculations
analogous to the ones discussed earlier give the following axion parameters
for the KSVZ model\cite{RDP}
\begin{equation}
\lambda_m = 1~; ~~~\lambda_3 = -\frac{1}{2}
\frac{(m_d-m_u)}{(m_d+m_u)}~; ~~~ \lambda_o = -\frac{1}{2}
\end{equation}
and
\begin{equation}
K_{a\gamma\gamma} = 3q_X^2 -
\frac{4m_d+m_u}{3(m_d+m_u)}~.
\end{equation}

\subsection*{DFSZ Axions}

In this class of models\cite{DFSZ}, both the quarks and leptons carry
$U(1)_{\rm PQ}$ and hence one again needs two Higgs fields $\Phi_1$ and
$\Phi_2$.  However, the quarks and leptons feel the effects of the axions
only through the interactions that the field $\sigma$ has with $\Phi_1$ and
$\Phi_2$ in the Higgs potential.  This interaction, conventionally, occurs
through the term
\begin{equation}
{\cal{L}}_{\rm axion} = \kappa\Phi_1^T C\Phi_2(\sigma^\dagger)^2 + {\rm h.c.}~,
\end{equation}
which serves to fix the PQ properties of $\sigma$ relative to 
$\Phi_1$ and $\Phi_2$.

The contribution of the axion field $a$ in $\Phi_1$ and $\Phi_2$ is readily
isolated\cite{RDP}.  Defining, as before, $v_1^2+v_2^2 = v^2$, one has
\begin{eqnarray}
\Phi_1 &=& \frac{v_1}{\sqrt{2}} \exp\left(i\frac{2v_2^2a}{v^2 f}\right)
\left[ \begin{array}{c} 1\\ 0 
\end{array} \right] \nonumber \\
&\equiv & \frac{v_1}{\sqrt{2}} \exp\left(i\frac{X_1}{f} a\right)
\left[ \begin{array}{c} 1 \\ 0
\end{array} \right] \\
\Phi_2 &=& \frac{v_2}{\sqrt{2}} \exp\left(i\frac{2v_1^2a}{v^2f}\right)
\left[ \begin{array}{c} 0 \\ 1
\end{array} \right] \nonumber \\
&\equiv & \frac{v_2}{\sqrt{2}} \exp\left(i\frac{X_2a}{f}\right)
\left[ \begin{array}{c} 0 \\ 1
\end{array} \right]~,
\end{eqnarray}
where $X_1 = 2v_2^2/v^2;~X_2 = 2v_1^2/v^2$.  These formulas correspond to
what was done earlier (cf. Eq. (39)) with the replacements:  $x\leftrightarrow
X_1$; $x^{-1} \leftrightarrow X_2$.  Using instead of $f$ the scale
\begin{equation}
\tilde f = f/2N_g
\end{equation}
in the formula for the axion mass and mixing parameters (Eqs. (35)-(37)) gives
the same expression for the axion mass in the DFSZ and KSVZ models.  
Adopting this convention, the axion parameters for the DFSZ case are
\begin{eqnarray}
\lambda ~&=& 1~; ~~~ \lambda_3 = \frac{1}{2}
\left[\frac{X_1-X_2}{2N_g}-\frac{(m_d-m_u)}{(m_d+m_u)}\right]~; \nonumber \\
\lambda_o &=& \frac{1-N_g}{2N_g} 
\end{eqnarray}
and
\begin{equation}
K_{a\gamma\gamma} = \frac{4}{3} - \frac{4m_d+m_u}{3(m_d+m_u)}~.
\end{equation}
Note that the DFSZ axion has a coupling to electrons given by
\begin{equation}
{\cal{L}}^{\rm DFSZ}_{\rm aee} = -i\frac{X_2}{2N_g}
\frac{m_e}{\tilde f} a\bar e\gamma_5 e~.
\end{equation}

\section*{Astrophysical and Cosmological Constraints on Invisible Axions}

Invisible axion models are constrained substantially by astrophysical and
cosmological considerations, which restrict the allowed range for
$f(\tilde f)$---or equivalently the range for the axion mass
\begin{equation}
m_a \simeq 6.3 \left[\frac{10^6 ~{\rm GeV}}{f}\right]~{\rm eV}~.
\end{equation}
The astrophysical bounds on $m_a$ arise, essentially, because axion emission
removes energy from stars altering their evolution.  Because the axion
couplings to matter are inversely proportional to $f$, for large values of
$f$ (and therefore small axion masses) axions are less effective at cooling
stars.  Hence astrophysics provides a lower bound for $f$ or, equivalently,
an upper bound on $m_a$.  If axions are sufficiently light, their emission
eventually is irrelevant for the evolution of stars.

The bounds on invisible axions one derives from stellar evolution
considerations are less restrictive for KSVZ axions.  It turns out that the
most effective cooling process is Compton axion production\cite{Raffelt}
$\gamma e\to ea$, which is proportiional to the coupling of axions to 
electrons.  This process is absent for KSVZ axions, since $g^{\rm KSVZ}_{eea}
=0$.  However, one can obtain a bound on these axions since they can remove
energy from stars via the Primakoff process $\gamma Z,e\to Z,ea$ involving the
axion to 2-photon coupling $K_{a\gamma\gamma}$.

Schematically, the energy lost by a star from the Compton process is given by
\begin{equation}
Q = \frac{1}{\rho_{\rm star}} \int dn_e dn_\gamma |v|\sigma E_a~,
\end{equation}
where $\rho_{\rm star}$ is the stellar density and one integrates the
interaction rate $|v|\sigma$ weighted by the axion energy over the density
of the initial states.  For $m_a\ll T_{\rm star} \sim O(KeV)$, the only
dependence on $m_a$ are through the dependence of the interaction rate on
$f$: $|v|\sigma \sim f^{-2}\sim m_a^2$.  Bounds are then obtained by
requiring that $Q\leq Q_{\rm nucl}$--the rate of nuclear energy 
generation ($Q_{\rm nucl} \sim 10^2 ~\hbox {ergs/g sec}$).

From a detailed and careful study of how axion emission would affect stellar
evolution, Raffelt\cite{Raffelt} gives the following bounds on the two
classes of invisible axions
\begin{equation}
(m_a)_{\rm DFSZ} \leq \frac{10^{-2}}{X_2} {\rm eV}~; ~~~
(m_a)_{\rm KSVZ} \leq \frac{0.27}{K_{a\gamma\gamma}} {\rm eV}~.
\end{equation}
Stronger bounds than these can be derived from the observation of neutrinos
from SN 1987a.  The bounds arise because if the axion luminosity is 
comparable to the neutrino luminosity ($\sim 10^{53}~{\rm ergs/sec}$)
during the core collapse, then the neutrino signal would be altered.  It
turns out that the dominant process for axion production during the
collapse is axion bremstrahlung off nucleons $(N+N\to N+N+a)$.  As a result
the SN 1987a bounds one obtains are quite similar for KSVZ and DFSZ axions.
The results that Turner\cite{Turner} gives from his study of this issue are
\begin{eqnarray}
(m_a)_{\rm DFSZ} &\leq &\frac{1.7\times 10^{-3}}{\xi^2_{\rm brem}} 
{\rm eV}~; \nonumber \\ 
(m_a)_{\rm KSVZ} &\leq & 8.4\times 10^{-4}~{\rm eV}
\end{eqnarray}
where
\begin{equation}
\xi^2_{\rm brem} = 1.44 + \frac{1}{2} (X_1-X_2 - 1.55)^2~.
\end{equation}

Cosmology, on the other hand, provides an upper bound for $f~(\tilde f)$ or,
equivalently, a lower bound for the axion mass.  This bound was derived by a
number of authors\cite{Cosbo} and its origin is easy to understand.  When the
Universe goes through the $U(1)_{\rm PQ}$ phase transition, at temperatures
of order $T\sim f$, the axion field acquires a vacuum expectation.  At this
stage the color anomaly is not effective, so the axion is a Nambu-Goldstone
boson and $\langle a_{\rm phys.}\rangle \sim f$.  As the Universe cools to a
temperature $T^*\sim \Lambda_{\rm QCD}$, the axion gets a mass of order
$m_a\sim \Lambda^2_{\rm QCD}/f$ and the axion VEV is driven to zero
dynamically $\langle a_{\rm phys.}\rangle \to 0$, corresponding to
$\bar\theta = 0$.  The relaxation of $\langle a_{\rm phys.}\rangle$ to this
value is oscillatory and this coherent oscillation of the zero-momentum
component of the axion field contributes to the Universe's energy density.
The larger $f$ is, the larger the axion contribution to the energy density of
the Universe is.  Asking that this contribution not exceed the Universe's
closure density then gives an upper bound on $f$:

A little more quantitatively\cite{RDP}, one can examine the equation of motion
for $\langle a_{\rm phys.}\rangle$ in the expanding Universe in the 
approximation that only the axion mass term is relevant in the axion potential:
\begin{equation}
\frac{d^2\langle a_{\rm phys.}\rangle}{dt^2} + 3\frac{\dot R(t)}{R(t)}
\frac{d\langle a_{\rm phys.}\rangle}{dt} + m_a^2(t)
\langle a_{\rm phys.}\rangle = 0~,
\end{equation}
where $R(t)$ is the cosmic scale parameter and $\dot R(t)/R(t) = H(t)$ is the
Hubble constant.  If $H(t) \gg m_a(t)$ then there are no oscillations, while
in the reverse limit the oscillations are sinusoidal.  Oscillations start
at $T^*$, when
\begin{equation}
m_a(T^*) \sim H(T^*) \sim \frac{\Lambda^2_{\rm QCD}}{M_{\rm Planck}}~.
\end{equation}
At the start of oscillations, the energy density in the axion field is of
order $\rho_a(T^*)\sim m_a^2(T^*)f^2$.  If one assumes that $m_a(t)$ is
slowly varying then one can show that\cite{Cosbo} $\rho_a(t)\sim m_a(t)/
R^3(t)$.  Thus the contribution of axion oscillations to the Universe energy
density today is of order
\begin{equation}
\rho_a = \rho_a(T^*)
\left[\frac{m_a}{m_a(T^*)}\right]\left[\frac{R^3(T^*)}{R^3}\right] \sim
\frac{\Lambda^3_{\rm QCD} T^3}{m_a M_{\rm Planck}}~,
\end{equation}
where $T\sim 3^\circ K$ is the temperature of the Universe today.  
Requiring that $\rho_a$ be less than the closure density of the Universe
provides the lower bound on $m_a$.  Amazingly, the above order of magnitude
formula gives the same bounds on $m_a$ that the more careful calculations 
give\cite{RDP}
\begin{equation}
m_a\geq (10^{-5}-10^{-6})~{\rm eV}~.
\end{equation}

\section*{Looking for Invisible Axions}

The cosmological and astrophysical constraints on invisible axions allow only
a rather narrow window for the axion mass, roughly
\begin{equation}
10^{-6}~{\rm eV} \leq m_a \leq 10^{-3}~{\rm eV}~.
\end{equation}
If the axion has a mass near the lower limit, then axions play a very important
role in the Universe and could be (part of) the Universe's cold dark matter.
The cosmological bounds are subject to caveats, so it is useful to explore all
masses below $10^{-3}~{\rm eV}$.  For instance, decay of axionic 
strings\cite{Davis} may increase the contribution of axions to the present
Universe's energy density by about a factor of $10^3$--thereby pushing the
axion mass lower bound to the astrophysical limit.  On the other hand,
Linde\cite{Linde} has argued that in inflationary models, it is quite possible
that $\langle a_{\rm phys}(f)\rangle\sim cf$, with $c\ll 1$.  This would
decrease the lower bound on the axion mass by a factor of $c^2$.

These uncertainties notwithstanding, it is very worthwhile trying to detect
a possible signal of invisible axions.  Indeed, experiments are presently
under way to try to detect such axions on the assumption that they constitute
the dark matter component of our galactic halo.  Under this assumption one can
estimate both the presumed axion energy density, $\rho^{\rm halo}_a \sim
5\times 10^{-25}\hbox{g cm}^{-3}\sim 300~{\rm MeV~cm}^{-3}$, and their
velocity, $v_a\simeq 10^{-3}$--the virial velocity in the galaxy.  The basic
idea for these experiments was suggested more than a decade ago by 
Sikivie\cite{Sikivie}.  It makes use of the axion to 2-photon coupling to 
resonantly convert in a laboratory magnetic field the halo axions into photons,
which can then be detected in a cavity.

The interaction of halo axions with a constant magnetic field $B_o$ in a 
resonant cavity, as a result of Eq. (37), produces an electric field with
frequency $\omega = m_a$.  The generated electromagnetic energy can be
detected in the cavity.  When the cavity is tuned to the axion frequency, one
should observe a narrow line on top of the noise spectrum.  On resonance, one
can write for the axion to photon conversion power the expression\cite{cavity}
\begin{equation}
P=\left[\frac{\rho_a^{\rm halo}}{m_a}\right][VB_o^2]
\left[\frac{\alpha}{\pi f} K_{a\gamma\gamma}\right]^2 C_{\rm over} Q_{\rm eff}~.
\end{equation}
Here the first factor is just the number of axions per unit volume; the second
details the magnetic energy; the third is the axion coupling strength
$(g^2_{a\gamma\gamma})$; the fourth is an overlap factor of $O(1)$ which
depends in detail on which mode is excited in the cavity.  Finally the last
factor is the effective $Q$ for the experiment, which is the minimum of the
cavity's $Q$ and that produced by the spread in the axion frequency due to its
velocity.  Typically both numbers are of $O(10^6)$.

Halo axions produce microwave photons $(1~{\rm GHz} = 4\times 10^{-6}~{\rm eV})$.
Two pilot experiments, done at Brookhaven\cite{BNL} and the University of
Florida\cite{UF} had limited ``magnetic energy"--typically $B_o^2V\sim
0.5~{\rm (Tesla)}^2m^3$--and relatively noisy amplifiers.  These experiments set
limits for $g^2_{a\gamma\gamma}$ about a factor of 100-1000 away from
theoretical expectations.  There are now, however, two second generation
experiments which should be sensitive at, or near, the theoretical limits.

%\begin{figure}
%\begin{center}
%test%~\epsfig{file=fig.ps}
%\end{center}
%\end{figure}

The first of these experiments, which has just started taking data, is being
carried out at Livermore National Lab\cite{LNL}.  It has a very large magnetic
volume $(B_oV^2\sim 12~{\rm (Tesla)}^2m^3)$ and state of the art
amplifiers to reduce the noise level\footnote{Even so the power produced by
halo axions is tiny.  At 1 GHz one expects $P_{\rm axion} \sim 4\times 10^{-20}~
{\rm Watts}$!}.  The second experiment, to be carried out in Kyoto 
University\cite{Kyoto}, uses a rather moderate magnetic volume
$(B_oV^2\sim 0.2~({\rm Tesla})^2m^3)$.  However, it employs an extremely
clever technique for counting the produced photons using Rydberg atoms, which is
extremely sensitive.  I show in the Figure the expectations of these experiments,
along with the limits obtained by the pilot experiments.  It is clear that 
these experiments have the capability to answer the question of the existence
of axions in the mass range that they can probe.  Results are expected in the
next few years.  So, stay tuned!

\section*{Acknowledgements}

I am grateful to the organizers of the KOSEF-JSPS Winter School for their
gracious hospitality in Seoul.  This work was supported in part by the
Department of Energy under Grant No. FG03-91ER40662.

%\section*{References}

\end{document}